\documentclass[showpacs,amsmath,amssymb,aps,10pt,reprint,superscriptaddress,prl]{revtex4-1}
\usepackage{graphicx}
\usepackage{bm}
\usepackage[breaklinks=true,colorlinks=true,linkcolor=blue,urlcolor=blue,citecolor=blue]{hyperref}
\usepackage{graphics}
\usepackage{dcolumn}
\usepackage{amsmath,amssymb}
\usepackage{mathdots}
\usepackage{natbib}
\usepackage{soul,color}
\usepackage{epstopdf}
\usepackage{wasysym}

\usepackage{lipsum}

\begin{document}
\abovedisplayskip=3pt
\belowdisplayskip=3pt
\abovedisplayshortskip=2pt
\belowdisplayshortskip=2pt


\title{Microwave emission from superconducting vortices in Mo/Si superlattices}

\author{O. V. Dobrovolskiy}
    \email{Dobrovolskiy@Physik.uni-frankfurt.de}
    \affiliation{Physikalisches Institut, Goethe University, 60438 Frankfurt am Main, Germany}
    \affiliation{Physics Department, V. Karazin Kharkiv National University, 61077 Kharkiv, Ukraine}
\author{V. M. Bevz}
    \affiliation{Physics Department, V. Karazin Kharkiv National University, 61077 Kharkiv, Ukraine}
\author{M.~Yu.~Mikhailov}
    \affiliation{B. I. Verkin Institute for Low Temperature Physics and Engineering, 61103 Kharkiv, Ukraine}
\author{O.~I.~Yuzephovich}
    \affiliation{B. I. Verkin Institute for Low Temperature Physics and Engineering, 61103 Kharkiv, Ukraine}
\author{V.~A.~Shklovskij}
    \affiliation{Physics Department, V. Karazin Kharkiv National University, 61077 Kharkiv, Ukraine}
\author{R.~V.~Vovk}
    \affiliation{Physics Department, V. Karazin Kharkiv National University, 61077 Kharkiv, Ukraine}
\author{M. I. Tsindlekht}
    \affiliation{The Racah Institute of Physics, The Hebrew University of Jerusalem, 91904 Jerusalem, Israel}
\author{R. Sachser}
    \affiliation{Physikalisches Institut, Goethe University, 60438 Frankfurt am Main, Germany}
\author{M. Huth}
    \affiliation{Physikalisches Institut, Goethe University, 60438 Frankfurt am Main, Germany}
\date{\today}

\begin{abstract}
Most of superconductors in a magnetic field are penetrated by a lattice of quantized flux vortices. In the presence of a transport current causing the vortices to cross sample edges, emission of electromagnetic waves is expected due to the continuity of tangential components of the fields at the surface. Yet, such a radiation has not been observed so far due to low radiated power levels and lacking coherence in the vortex motion. Here, we report emission of electromagnetic waves from vortices crossing the layers of a superconductor/insulator Mo/Si superlattice. The emission spectra consist of narrow harmonically related peaks which can be finely tuned in the GHz range by the dc bias current and, coarsely, by the in-plane magnetic field value. Our findings show that superconductor/insulator superlattices can act as dc-tunable microwave generators bridging the frequency gap between conventional radiofrequency oscillators and (sub-)terahertz generators relying upon the Josephson effect.
\end{abstract}

\keywords{Abrikosov vortices, vortex dynamics, Mo/Si superlattice, matching field, electromagnetic radiation}
\maketitle
\clearpage

In 1936, Shubnikov \emph{et al.} \cite{Sch36pzs,Shu08ujp} discovered a new type of superconductors, now called type-II superconductors, which are in the core of modern superconducting technology. As distinct from type-I superconductors which carry supercurrents at the surface and whose superconducting state is destroyed at relatively weak fields, type-II superconductors are capable of carrying bulk supercurrents and maintain a very-low-dissipative state up to high magnetic fields. The understanding of the Shubnikov phase, referred to as a mixed state of type-II superconductors, came after the Nobel Prize work by Abrikosov \cite{Abr57etp} who elucidated that a magnetic field, whose magnitude is between the lower and upper critical field, penetrates type-II superconductors as a lattice of quantized magnetic vortices. Each vortex carries one quantum of magnetic flux $\Phi_0 = 2.07\times10^{-15}$\,Vs and can be regarded as a tiny whirl of the supercurrent. When a rather large external current is applied to a type-II superconductor, the vortices move under the action of the Lorentz force and this vortex movement is accompanied by oscillations of the supercurrents and the associated magnetic induction \cite{Kul66spj,Sch66phl}. Thus, supercurrent oscillations arising from vortex motion have been experimentally observed in granular superconducting films by Martinoli \emph{et al.} \cite{Mar76prl} and Hebboul \emph{et al.} \cite{Heb99prl}. As the vortex lattice comes to a sample edge, the oscillating electric and magnetic fields of vortices should propagate into free space due to the continuity of tangential components of the fields at the surfaces \cite{Dol00prb,Bul06prl}. The spectrum of the electromagnetic (em) radiation from the vortex lattice crossing a sample edge has been predicted to peak at the harmonics of the washboard frequency $f_0 = v/d$, where $v$ is the vortex velocity and $d$ is the distance between the vortex rows in the direction of motion. Yet, such a radiation has not been observed so far, as its detection poses a severe experimental challenge. Namely, while the radiated power from a 1\,mm$^2$ sample surface has been estimated to be of the order of $\sim10^{-7}$\,W for a triangular vortex lattice, this value drastically decreases in the presence of disorder \cite{Bul06prl}. While the vortex flow is known to become unstable at vortex velocities of the order of $1$\,km/s \cite{Lar75etp,Shk17prb}, for an em generation at $f_0 \simeq 10$\,GHz, i.e. in the frequency range which is important for signal processing, modulation of the em properties of the superconductor at a length scale of and below $100$\,nm is required. Unfortunately, a thin film geometry with an out-of-plane magnetic field makes an em generation detection barely feasible, as the area of the side surfaces crossed by vortices becomes negligibly small. In addition, uncorrelated disorder reduces the range of correlations in the vortex lattice, thus suppressing the radiated power levels even further.

Here, we provide experimental evidence for the em radiation from a lattice of Abrikosov vortices moving across the layers in a superconductor/insulator Mo/Si superlattice. Emission powers at levels above $10^{-12}$\,W are observed for a $5$\,mm$^2$ sample surface crossed by vortices in the coherent regime achieved at large matching values  of the magnetic field when a dense vortex lattice is commensurate with the multilayer period. The emission is peaked at the harmonics of the washboard frequency $f_0$ which can be finely tuned from about $5$\,GHz to about $30\,$GHz by the dc bias current and, coarsely, by switching the in-plane magnetic field between matching values. Furthermore, by varying the size of the vortex cores by temperature, we exploit the dimensionality crossover of superconductivity in the superlattice for tailoring the emission spectra. Namely, we tune the frequency-selective em emission at the harmonics of $f_0$ related to the period of the vortex lattice crossing the sample edges to the harmonics of $2f_0$ related to the multilayer period when vortices fit in the insulating Si layers. Our findings show that superconductor/insulator multilayers can act as dc-tunable microwave generators bridging the frequency gap between conventional radiofrequency oscillators and (sub-)terahertz generators relying upon the Josephson effect.

\section{Results}
\begin{figure}[b!]
\centering
    \includegraphics[width=1\linewidth]{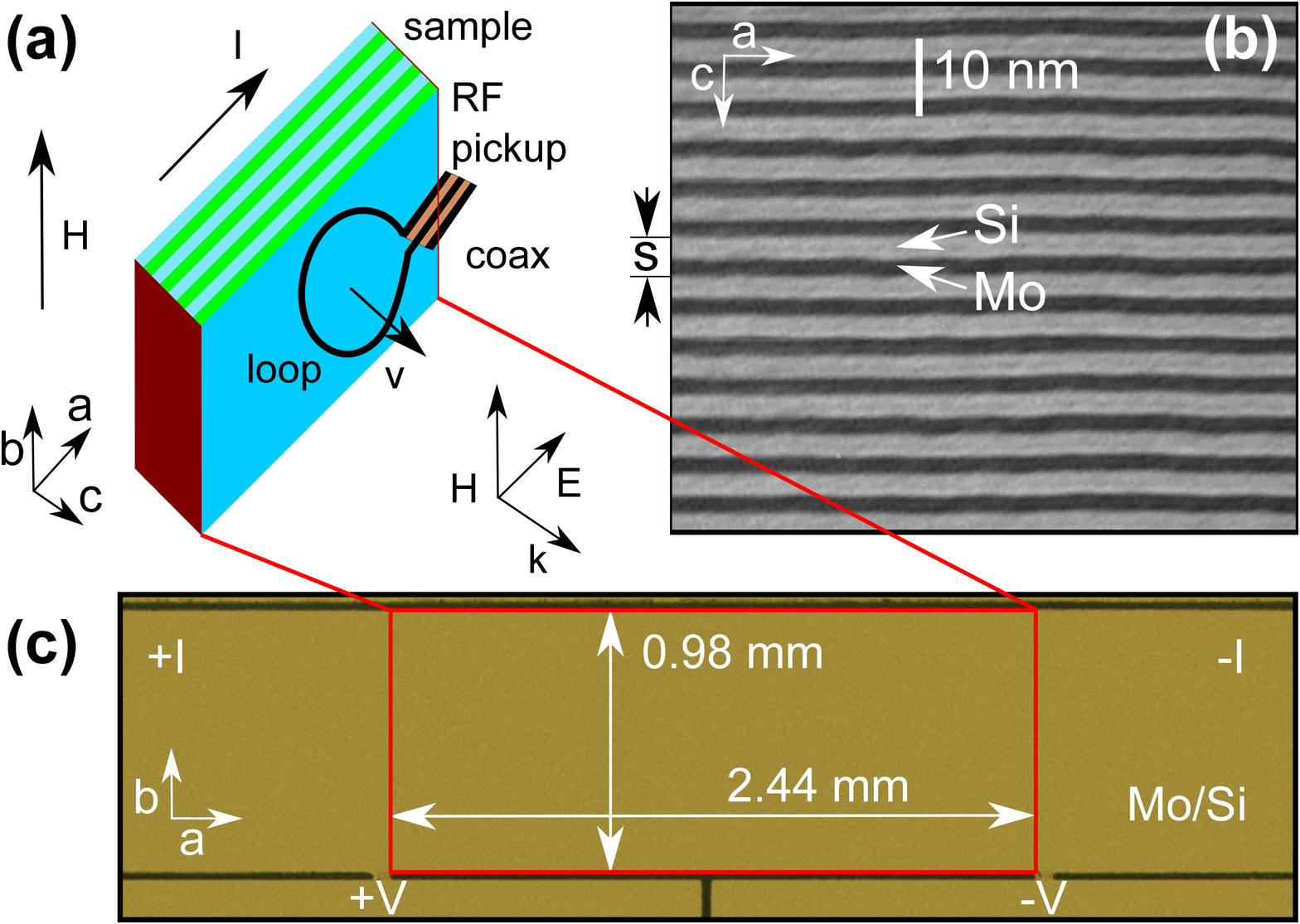}
    \caption{\textbf{Superconducting/insulator Mo/Si superlattice.}
    (\textbf{a}) Experimental geometry (not to scale). The Mo/Si multilayer is in a magnetic field $H$ applied parallel to the $b$-axis. The transport current $I$ applied along the $a$ axis causes the vortex lattice to move with velocity $v$ across the layers. Electromagnetic radiation from the lattice of flux lines crossing the superconducting layers is picked-up by a wire loop antenna. See Methods for details.
    (\textbf{b}) Transmission electron microscopy image of the Mo/Si multilayer with a multilayer period $s=d_\mathrm{Mo} +d_\mathrm{Si} =50\,\mathrm{\AA}$.
    (\textbf{c}) Optical microscope image of the bridge etched in the Mo/Si multilayer.}
   \label{f1}
\end{figure}

\textbf{Investigated system.} The investigated system is shown in Fig.\,\ref{f1}(a). The emission of em waves at microwave frequencies is detected from Abrikosov vortices crossing the layers in a superconductor/insulator Mo/Si superlattice. The superlattice consists of $50$ alternatingly sputtered Mo and Si layers with thicknesses $d_\mathrm{Mo}=22\,\mathrm{\AA}$ and $d_\mathrm{Si} = 28\,\mathrm{\AA}$, resulting in a multilayer period $s$ of $50\,\mathrm{\AA}$. A transmission electron microscopy image of a part of the sample is shown in Fig. \ref{f1}(b). The superconducting transition temperature of the sample, determined at the midpoint of the resistive transition $R(T)$, is $T_c = 4.02$\,K. The Josephson coupling between the superconducting Mo layers is rather strong $\eta_J = \hbar^2/2m^2s^2 \gamma \approx1$ \cite{Mik99ltp}. A four-probe $5\times1\,$mm$^2$ bridge was patterned in the sample for electrical transport measurements, Fig. \ref{f1}(c). The magnetic field and transport current were applied in the layer plane and orthogonal to each other, causing a vortex motion across the layers under the action of the Lorentz force, Fig. \ref{f1}(a). The emitted signal was picked up by a small wire loop shorting the end of a semirigid coaxial cable and placed close to the sample surface \cite{Gol94prb}. The em emission was monitored by a high-frequency spectrum analyzer.

\begin{figure*}[t!]
\centering
    \includegraphics[width=1\linewidth]{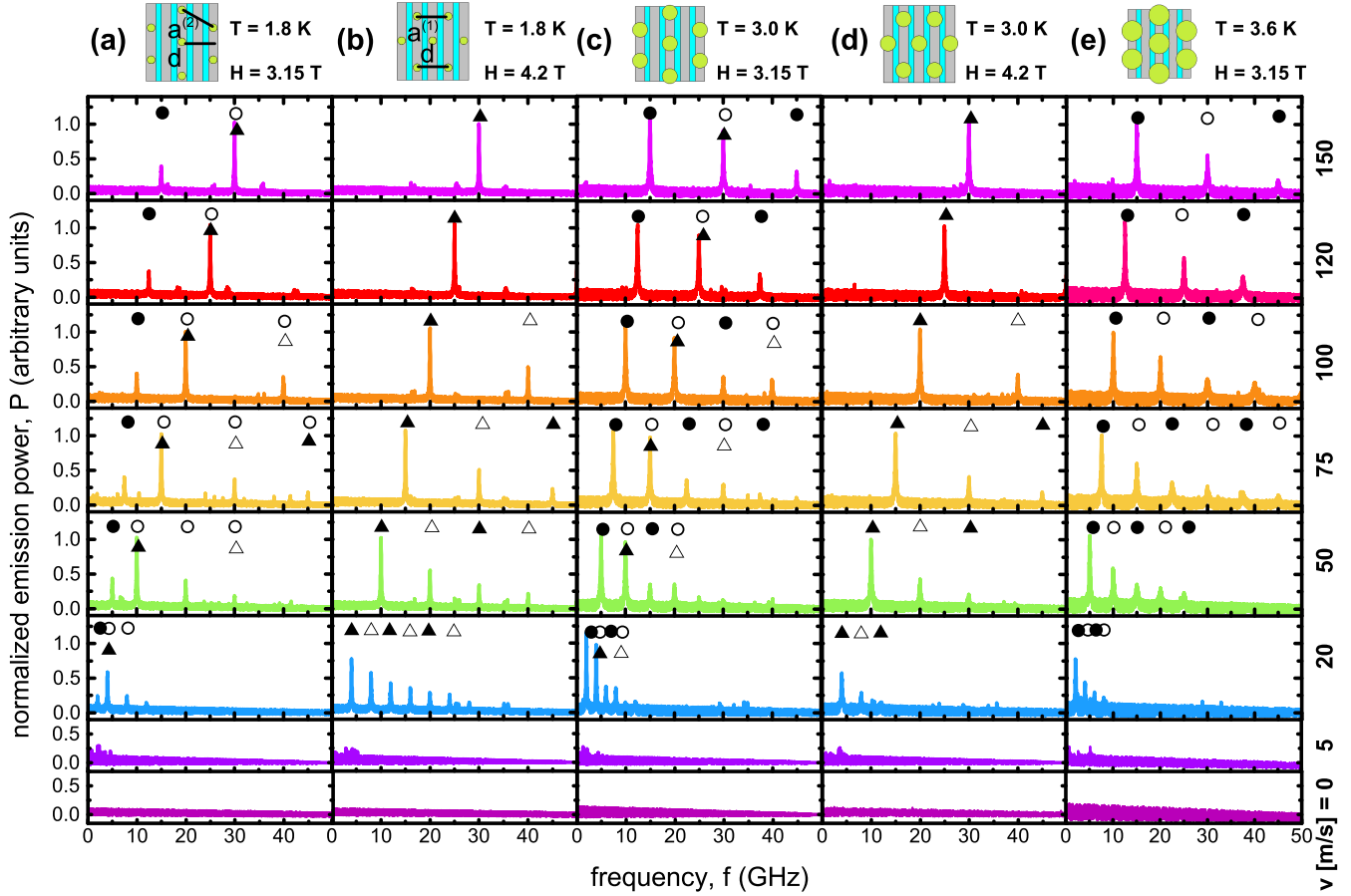}
    \caption{\textbf{Microwave emission from Abrikosov vortices in Mo/Si superlattices.}
    Emission spectra for a series of vortex velocities $v$, as indicated, at $H = 3.15$\,T, $4.2$\,T and $T = 1.8$\,K, $3.0$\,K, and $3.6$\,K. The symbols above the emission peaks indicate the frequencies $f_m = m f_0$ which are harmonically related to the washboard frequency $f^{(1)}_0$ associated with the $50\,\mathrm{\AA}$-periodic layered structure (triangles) and $f^{(2)}_0$ related to the $100\,\mathrm{\AA}$-spaced vortex rows in the direction of their motion (circles). The odd and even harmonics are indicated by solid and open symbols, respectively. The vortex lattice configurations, which are commensurate with the Mo/Si multilayer period at $H^{(2)}_{N=2} = 3.15$\,T with $a^{(2)} = 2d/\sqrt{3} =4s/\sqrt{3}$ and $H^{(1)}_{N=1} = 4.2$\,T with $a^{(1)} = d = 2s$, are shown in the scaled coordinate system $(\gamma a,c)$ above the panels. The size of the circles denoting the vortex cores (not to scale) reflects their relation to the Si layer thickness $d_{\mathrm{Si}}$ and the multilayer period $s$.}
   \label{f2}
\end{figure*}

\begin{figure*}[t!]
\centering
    \includegraphics[width=0.75\linewidth]{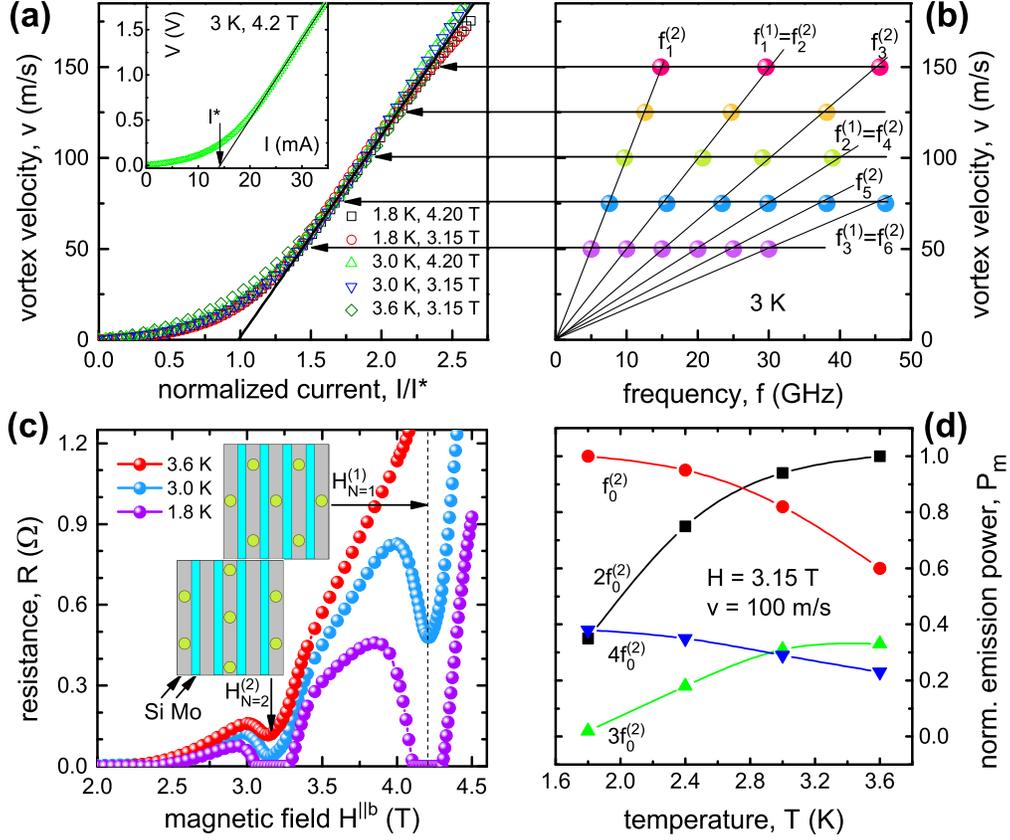}
    \caption{\textbf{Emission frequencies and vortex lattice configurations at matching fields.}
    (\textbf{a}) $I$-$V$ curves in the vortex velocity versus normalized current representation. The horizontal arrows indicate the vortex velocities at which the emission spectra in Fig.\,\ref{f2} have been acquired. Inset: The $I$-$V$ curve of the sample at 3\,K and 4.2\,T. The vertical arrow indicates the definition of $I^\ast$ used for plotting the $I$-$V$ curves in the main panel.
    (\textbf{b}) Peak frequencies versus vortex velocity for the data sets (c) and (d) of Fig.\,\ref{f2}.
    (\textbf{c}) Resistance as a function of $H^{\parallel b}$ for a series of temperatures, as indicated. The vortex lattice configurations, which commensurate with the Mo/Si superlattice at $H^{(2)}_{N=2} = 3.15$\,T and $H^{(1)}_{N=1} = 4.2$\,T, are shown in the scaled coordinate system $(\gamma a,c)$ in the inset.
    (\textbf{d}) Normalized emission power $P_m$ as a function of temperature for the first four lowest-order harmonics $f^{(2)}_0=9.98\,$GHz emitted at the vortex velocity $v=100$\,m/s at $3.15$\,T. Sold lines are guides for the eye.
    }
   \label{f3}
\end{figure*}

\textbf{Microwave emission from moving vortex lattice.} Figure \ref{f2} displays the emission spectra recorded at vortex velocities $v$ from $0$ to $150$\,m/s in magnetic fields $H = 3.15$\,T and $4.2$\,T at temperatures $T = 1.8$\,K, $3$\,K, and $3.6$\,K. At $v = 0$ the noise floor of the detector is seen and there is no emission detected. At currents exceeding the depinning current the vortices move across the layers. Since the depinning current decreases with increase of $T$ and $H$, in Fig. \ref{f3}(a) we plot the $I$-$V$ curves in a vortex velocity $v$ versus normalized current $I/I^\ast$ representation. Here, the vortex velocity $v$ was deduced from the $I$-$V$ curves using the standard relation $v = V/HL$, where $V$ is the measured voltage, $H$ is the magnetic field value and $L = 2.44$\,mm is the distance between the voltage contacts. The $I^\ast$ values were deduced for each of the $I$-$V$ curves at the intersect of the extrapolated linear section with the current axis, as exemplified in the inset of Fig. \ref{f3}(a). Accordingly, $I^\ast$ has the meaning of a depinning current determined by the ``dynamic'' criterion, as is commonly used for systems with strong pinning \cite{Bla94rmp}. For vortex velocities between $50$\,m/s and $150$\,m/s the scaled $I$-$V$ curves in Fig. \ref{f3}(a) fit to the universal relation $v = 109.5(1- I/I^\ast)\,\mathrm{m/s}$ with an error of less than 5\,\%. This relation not only allows for a direct comparison of the emission spectra acquired at different $T$ and $H$ values in Fig. \ref{f2} using $v$ as a deduced parameter, but it also links the peak frequencies with the dc bias current $I$ which is a driving parameter in our experiment. In particular, with increase of the vortex velocity to above $20$\,m/s a series of peaks appears in all panels of Fig. \ref{f2} on the background of the noise floor. The peaks are best seen in the range of vortex velocities between $50$\,m/s and $150$\,m/s corresponding to the nearly linear regime of viscous flux flow in Fig. \ref{f3}(a). We note that the higher-frequency peaks at $f_m = mf_0$ are harmonically related to the lowest-frequency peak at $f_0$. The largest number of harmonics $m =6$ is observed in the accessible frequency range at a vortex velocity $v = 75$\,m/s. Except for the data set (a) in Fig. \ref{f2}, to which we return in what follows, the peak power $P_m$ decreases with increasing $f$. We emphasize, that whereas $f_0$ does not depend on temperature, it does depend on the magnetic field. This is why in what follows we will distinguish between $f^{(2)}_0$ for the data sets (a), (c) and (e) acquired at $H = 3.15$\,T and $f^{(1)}_0$ for the data sets (b) and (d) for $H = 4.2$\,T. Importantly, $f^{(1,2)}_0$ are shifted towards higher frequencies with increase of the dc bias current. Specifically, $f^{(2)}_0 = 5.01\,$GHz at $3.15$\,T and $f^{(1)}_0 = 9.98\,$GHz at $4.2$\,T at $v = 50$\,m/s evolve into $f^{(2)}_0 = 15.04\,$GHz and $f^{(1)}_0 = 29.87\,$GHz at $v = 150$\,m/s, respectively. A linear dependence of the peak frequencies on the vortex velocity becomes apparent in Fig. \ref{f3}(b) where the data deduced from panel columns (c) and (d) of Fig. \ref{f2} are presented. Evidently, the observed emission is related to the washboard frequency associated with the vortex dynamics.

\textbf{Vortex lattice configurations at matching fields.} The $H$ values at which the spectra in Fig. \ref{f2} have been acquired correspond to minima in the resistance curves $R(H^{\parallel b})$ shown in Fig. \ref{f3}(c). Namely, at $T = 3.6$\,K, $R(H^{\parallel b})$ has a minimum centered at $H = 3.15$\,T.  At $T = 3$\,K the minimum at $H = 3.15$\,T becomes deeper and a second minimum appears at $H = 4.2$\,T. At $T = 1.8$\,K the two minima evolve into zero-resistance states in fields $3-3.3$\,T and $4.15-4.35$\,T. For the elucidation of what periodic length scale in the studied system is associated with the peaks at the different $H$, we analyze the stable vortex lattice configurations at the resistance minima in Fig. \ref{f3}(c) in Supplementary Materials and just outline the results of this analysis here. Namely, the commensurability effect in anisotropic layered superconductors was considered theoretically by Bulaevskij and Clem (BC) \cite{Bul91prb} on the basis of the discrete Lawrence-Doniach approach and by Ivlev, Kopnin, and Pokrovsky (IKP) in the framework of the continuous Ginzburg-Landau model \cite{Ivl90ltp}. The $R(H^{\parallel b})$ curve of our sample has no minima at the BC matching fields. We attribute this to a relatively large interlayer coupling in our sample and compare the data with the continuous IKP model. Namely, the IKP matching fields in our data range are $H^{(1)}_{N=1} = 4.2$\,T and $H^{(2)}_{N=2} = 3.15$\,T, in perfect agreement with the field values at which the resistance minima are observed in Fig. \ref{f3}(c). Accordingly, our analysis of the resistance minima suggests that we deal with a lattice of Abrikosov rather than Josepshon vortices. At the same time, we can not rule out a crossover from Abrikosov to Josephson vortices with further decrease of the temperature, as such a crossover is known in layered systems when the Abrikosov vortex with a suppressed order parameter in its core turns into a Josephson phase vortex once its core completely fits into the insulating layer \cite{Mol12nam}. Further support in favor of dealing with Abrikosov vortices is provided by the $I$-$V$ curves allowing for a universal scaling in the flux-flow regime, which would be impossible due to a sudden dissipation reduction at the crossover from Abrikosov to Josephson vortices \cite{Mol12nam}.

\section{Discussion}

\textbf{Superconductivity dimensionality crossover in the Mo/Si superlattice.} The evolution of the matching minimum in the $R(T)$ curve at $T = 3.6$\,K to the zero-resistance state at $T = 1.8$\,K in Fig. \ref{f3}(c) can be understood with the aid of the superconductivity dimensionality crossover occurring in the Mo/Si superlattice, as inferred from the $H$-$T$ phase diagram shown in Fig. 1s(a) in the Supplementary Materials. The out-of-plane upper critical field extrapolated to zero temperature $H_{c2}^{\parallel c}(0) = 8.4$\,T yields $\xi_{ab} = [\Phi_0/2\pi H_{c2}^{\parallel c} (0)]^{1/2} = 63\,\mathrm{\AA}$ and, hence, $\xi_c(0) = \xi_{ab}(0)/\gamma = 12\,\mathrm{\AA}$. In the $H$-$T$ diagram, there is a crossover temperature $T^\ast = T_c(1-\tau)$ with $\tau = 2\xi^2_c/s^2 \approx 3.60$\,K below which the system behaves in a 2D manner and exhibits a 3D behavior at $T>T^\ast$. The increase of the size of the vortex core with increasing temperature $\simeq2\xi_c(T)$ is illustrated in Fig. 1S in the Supplementary Materials in comparison with the thickness of the Si layer $d_{\mathrm{Si}}$ and the multilayer period $s$. As a brief summary of the analysis, a semi-quantitative relation of the vortex core size to the Si layer thickness and the multilayer period is sketched on the top of the spectra in Fig. \ref{f2}. In particular, at $1.8$\,K, being the lowest temperature accessible in our experiment, the vortex core $2\xi_c(1.8\,\mathrm{K}) \approx d_\mathrm{Si}=28\,\mathrm{\AA}$ largely fits into the insulating layers, thereby allowing the Mo layers to remain superconducting up to very high fields \cite{Ber97prl,Iid13prb}. At $3$\,K the vortex core $2\xi_c(3\,\mathrm{K}) \approx d_\mathrm{Si}\approx 50\,\mathrm{\AA}$ becomes comparable with the multilayer period. Even though some part of the vortices penetrates into the Mo layers, there are field ranges where the intrinsic pinning energy $E_p$ is larger than the elastic energy of a vortex lattice shear deformation $E_{el}$, which explains the presence of a rather broad resistance minimum in the vicinity of the matching fields. At $3.6$\,K the vortex cores become appreciably larger than the multilayer period, namely $2\xi_c(3.6\,\mathrm{K}) > 70\,\mathrm{\AA}$, such that the intrinsic confinement potential is smoothed out as the vortex core extends over more than one multilayer period. In this case the superlattice is no longer felt by a vortex as a layered structure, but rather the motion of vortices occurs in some effective continuous medium. Accordingly, the matching minimum at $3.15$\,T becomes shallow at $3.6$\,K while the minimum at $4.2$\,T disappears altogether as this field value is too close to $H_{c2}^{\parallel b}(\mathrm{3.6\,K}) = 5.2$\,T and it gets smeared by the transition to the normal state.

\textbf{Modification of the emission spectra due to the superconductivity dimensionality crossover.}
The relation of the vortex core size $\sim2\xi(T)$ to $d_{\mathrm{Si}}$ and $s$, as discussed above, allows for the following explanation of the differences in the microwave emission spectra in Fig. \ref{f2}. Each of the superconducting layers in the sample acts as an emitter of em waves, provided the vortex cores are not too large to move in a smoothed out periodic potential. By contrast, when the vortex core becomes larger than the multilayer period, the emission of em waves takes place at the sample surfaces, just as considered theoretically by Bulaevskii and Chudnovsky \cite{Bul06prl}. Indeed, in this case we observe an emission of em waves at the harmonics of the washboard frequency $f^{(2)}_m = mf^{(2)}_0 = mv/d$ with $d = 102\,\mathrm{\AA}$ nicely corresponding to the matching condition $2s = d = a^{(2)}\sqrt{3}/2$ for the triangular flux lattice with the parameter $a^{(2)} = (2\Phi_0/\sqrt{3}H)^{1/2} = 115.5\,\mathrm{\AA}$ at $H^{(2)}_{N=2} = 3.15$\,T, Fig. \ref{f2}(e). By contrast, in the case of vortices whose diameters are smaller than the multilayer period this ``surface-related'' microwave emission is superimposed with the em emission from individual layers. In particular, the data at $T = 1.8$\,K in Fig. \ref{f2}(b) and (d) corroborate that the emission of em waves becomes also possible at harmonics of $f^{(1)}_m = mf^{(1)}_0 = mv/d$ with $d = 49.2\,\mathrm{\AA}$ corresponding to $2s = d = a^{(1)} = (2\Phi_0/\sqrt{3}H)^{1/2} = 50\,\mathrm{\AA}$ at $H^{(1)}_{N=1} = 4.2$\,T. At the same time, the em emission peaked at halved frequency $f^{(2)}_m = mf^{(2)}_0$ and, hence, related to a doubled periodic length scale as compared to $f^{(1)}_0$ is clearly distinguishable at $H^{(2)}_{N=2} = 3.15$\,T, Fig. \ref{f2}(a) and (c). This suggests an interference of the emissions with the fundamental frequencies $f^{(2)}_0$ and $f^{(1)}_0$ in the resulting spectrum. Finally, when the vortex cores become larger than $d_{\mathrm{Si}}$ (and especially larger than $s$) the softening of the spatial profile of the order parameter results in a less efficient emission from individual layers, so that the emission from the sample surfaces starts to dominate the interlayer emission, Fig. \ref{f2}(e).

\begin{figure}[t!]
\centering
    \includegraphics[width=0.9\linewidth]{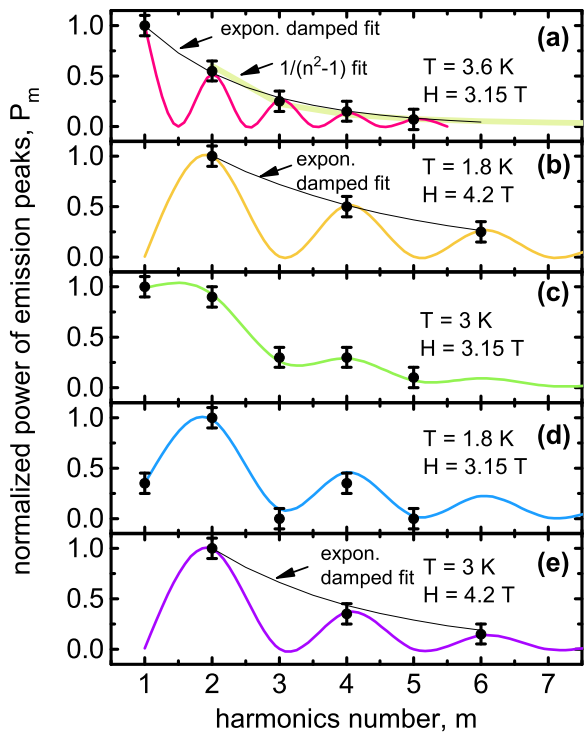}
    \caption{Normalized emission power $P_m$ as a function of the harmonics number $m$ for a vortex velocity of $v = 100$\,m/s. The assumed superposition of the em wave emission associated with the $100\,\mathrm{\AA}$-spacing between the vortex rows (a) and the $50\,\mathrm{\AA}$-periodic layered structure (b) is used to explain the spectrum modification in (c) and (d). The spectrum at $3$\,K (e) exhibits a stronger damping of higher harmonics as compared to $1.8$\,K (b). Symbols are $P_m$ values deduced from the spectra in Fig. \ref{f2}. Solid lines are fits as indicated.
    }
   \label{f4}
\end{figure}

\textbf{Interference of the interlayer- and surface-related emissions.} To support the assumption that in the general case the emission can be presented as a superposition of the emission at the sample surfaces and the interlayer emission, in Fig. \ref{f2} we denote the odd harmonics of $f^{(1)}_0$ with $m^{(1)}=1,3,5$ by $\blacktriangle$, the even ones with $m^{(1)}=2,4,6$ by $\vartriangle$, and the odd and even harmonics of $f^{(2)}_0$ with $\CIRCLE$ and $\Circle$, respectively. The normalized power $P_m$ of the emitted harmonics as a function of the harmonics number $m$ is plotted in Fig. \ref{f4}. We note that $P_m(m)$ follows an exponential decay  for the assumed dominating emission related to the $100\,\mathrm{\AA}$-periodic vortex row spacing in Fig. \ref{f4}(a) as well as for the assumed dominating emission related to the $50\,\mathrm{\AA}$-periodic superlattice in Fig. \ref{f4}(b) and (e). By contrast, the patterns of $P_m(m)$ at $3.15$\,T at lower temperatures are non-monotonic, Fig. \ref{f4}(c) and (d). We note that a decay of higher harmonics following the law $1/(n^2-1)$, which is a very good approximation to the exponentially damped fit for $n\leq 2$, was observed for the harmonics generation upon microwave transmission through granular YBCO thin films in the presence of an ac magnetic field \cite{Gol91prb}. Interestingly, an attenuation of higher harmonics in the electric field response in superconductors with a washboard pinning potential in the presence of a combination of dc and ac currents has been predicted to follow the envelope of modified Bessel functions \cite{Shk11prb}. Due to the mathematical analogy of the Langevin equation of motion of an Abrikosov vortex with the equation for the phase change in a Josephson contact the microwave Shapiro step amplitude follows the same law as a function of the microwave voltage in small junctions \cite{Gra81boo}.

\textbf{Frequency-selective generation by tuning the vortex core size.} If we now treat the exponentially damped curves in Fig. \ref{f4}(a) and (b) as functions enveloping harmonic functions with the period $p = m$ for the emission related to the $100\,\mathrm{\AA}$-periodic length scale in (a) and $p = 2m$ for the $50\,\mathrm{\AA}$-periodic length scale in (b), the $P_m$ pattern in panel (c) fits to a superposition of $0.32/0.68$-weighted functions from panels (a) and (b), while a good fit for $P_m$ in panel (d) is found for a superposition of $0.69/0.31$-weighted functions from panels (a) and (b). This allows for treating the $P_m(m)$ dependences in panels (c) and (d) as beatings of the $f_0$- and $2f_0$-waves, that suggests an incoherent interference of the microwave emission related to the $50\,\mathrm{\AA}$- and $100\,\mathrm{\AA}$-periodic length scales. In particular, the assumed beating of the $f_0$- and $2f_0$-waves allows us to explain the absence of peaks at $m=3,5$ at $T = 1.8$\,K in the data set (a) in Fig.\,\ref{f2}. The gradual decrease of the odd harmonics of $f^{(2)}_0$ accompanied by a simultaneous growth of the even ones at decreasing temperature becomes apparent in Fig. \ref{f3}(d). Thus, a suitable choice of the temperature allows for a frequency-selective generation of em waves evolving from the high-temperature spectrum containing a series of exponentially damped higher harmonics over an intermediate-temperature spectrum with higher harmonics obeying a more complex, beating-related law to the low-temperature spectrum in which the higher-order odd harmonics are absent. Furthermore, a faster attenuation of $P_m$ at $3\,$K in Fig. \ref{f4}(d) as compared to $1.8$\,K in Fig. \ref{f4}(b) can be attributed to softening of the spatial profile of the order parameter whose Fourier transform contains a smaller number of higher harmonics. Finally, the strongly suppressed em radiation at vortex velocities below $5$\,m/s might indicate that the typical time $t = 1/f = \Delta d/v \gtrsim 1\times10^{-11}$\,s of restoring the superconducting condensate upon crossing the edges ($\Delta d\sim5\,\mathrm{\AA}$) of the superconducting layers by vortices becomes sufficiently larger than the quasiparticles relaxation time in the studied system. This means that the variation of the magnetic induction as the vortices leave and enter the superconducting layers occurs adiabatically that can explain the absence of an emission in this quasistatic regime.

As an implication for superconducting applications, which can be drawn from our study, superconductor/insulator multilayers posses a potential for the use as on-chip generators. Their emission frequency $f_m = mf_0 = mv/d$ with $d = s$ (or $d=2s$, depending on the magnetic field value) can be monitored via the voltage drop related to the vortex velocity $v$ and finely tuned by the transport current, which is a driving parameter, via the relation $v = 109.5(1- I/I^\ast)\,\mathrm{m/s}$. The in-plane layout of Mo/Si superlattices allows for their on-chip integration with other fluxonic devices, such as diodes \cite{Vil03sci}, microwave filters \cite{Dob15apl} and transistors \cite{Vla16nsr} operating with Abrikosov vortices as well as quantum devices exploiting Josephson vortices as building blocks for coherent terahertz generation \cite{Wel13nph} and qubits for quantum computing \cite{Dev13sci}.

\textbf{Conclusion.}
To summarize, we have observed microwave radiation from a lattice of Abrikosov vortices moving across the layers in a Mo/Si superlattice. The emission spectrum is peaked at the harmonics of the washboard frequency $f^{(1)}_0$ related to the multilayer period and $f^{(2)}_0$ associated with the distance between the vortex rows in the direction of motion. The emission spectrum can be finely tuned by the dc bias current and, coarsely, by switching the in-plane magnetic field between matching values. In addition, we have revealed that the emission spectrum evolves as a function of temperature, such that the odd harmonics of the washboard frequency related to the distance between the vortex rows can be almost completely suppressed by choosing the matching field at which the vortex lattice is pinned in all neighboring insulating layers at lower temperatures. In all, our findings suggest that superconductor/insulator superlattices can act as dc-tunable microwave generators bridging the frequency gap between conventional rf oscillators and (sub-)terahertz generators relying upon the Josephson effect.

\clearpage
\section{Methods}
\textbf{Fabrication and properties of the Mo/Si superlattice.}
The superconductor/insulator superlattice consists of $50$ Mo and Si bilayers alternately sputtered onto a glass substrate at a substrate temperature of $100^\circ$C. The deposition rate was $2\,\mathrm{\AA}$/s. The thicknesses of the amorphous Mo and Si layers are $d_\mathrm{Mo}=22\,\mathrm{\AA}$ and $d_\mathrm{Si} = 28\,\mathrm{\AA}$, resulting in a superconducting layer repeat distance $s$ of $50\,\mathrm{\AA}$, which is referred to as a multilayer period. The individual layer thicknesses were inferred from small-angle x-ray reflectivity with an accuracy of $0.1\,\mathrm{\AA}$. The sample has 10\,nm-thick top and bottom Si layers. Its superconducting transition temperature, determined at the midpoint of the resistive transition $R(T)$, is $T_c = 4.02$\,K. This is noticeably higher than $T_c = 0.92$\,K of bulk Mo because of oscillations of $T_c$ of Mo/Si multilayers with increasing $d_\mathrm{Si}$ and an eventual saturation at $T_c = 7$\,K for $d_\mathrm{Si}> 120\,\mathrm{\AA}$ \cite{Fog96prb}. The interlayer Josephson coupling in the Mo/Si superlattice studied here is rather strong $\eta_J = \hbar^2/2m^2s^2 \gamma \approx1$ \cite{Mik99ltp}. A large ratio of the effective mass of the Cooper pairs $M$ perpendicular to the layer planes to the in-plane mass $m$ gives rise to an anisotropy $\gamma = (M/m)^{1/2}\approx5.22$ of the physical parameters of the superlattice, as inferred from the $H$-$T$ phase diagram shown in Fig. 1S(a) in the Supplementary Materials. These parameters include the in-plane ($ab$) and out-of-plane ($c$) upper critical field $H_{c2}^{\parallel ab} = \gamma H_{c2}^{\parallel c}$, the coherence length $\xi_{ab} =\gamma \xi_{c}$, and the penetration depth $\lambda_c = \gamma\lambda_{ab}$. The structure of individual vortices and the vortex lattice in the sample differs in essential ways from the conventional triangular vortex lattice in homogenous isotropic superconductors \cite{Bla94rmp}. Namely, the vortex core is elongated in the layer planes and compressed along the $c$-axis. For a magnetic field applied along the $b$-axis, the ground state vortex lattice configuration is given by a regular triangular lattice in the scaled coordinates $(a\gamma, c)$ \cite{Ivl90ltp,Kos13etp} which are used in all sketches in Fig. \ref{f2} and in the inset of Fig. \ref{f3}(c). A low-bound estimate for the zero-temperature gap frequency $2\Delta_0/h$ of the studied Mo/Si sample can be done using the standard BCS weak-coupling relation $\Delta_0 = 1.76k_B T_c$, which yields $f_G(0) \simeq300$\,GHz and $f_G \simeq100$\,GHz at $3.6$\,K, i.e. well above the highest frequency accessible in our experiment.

\textbf{Ultra-wide-band cryogenic spectroscopy.}
A four-probe $5\,$mm$\times1\,$mm bridge was patterned in the sample for electrical transport measurements. The distance between voltage contacts amounted to $2.44\,$mm. The magnetic field and transport current were applied in the layer plane and orthogonal to each other, causing a vortex motion across the layers under the action of the Lorentz force. The measurements were performed in a $^4$He cryostat with a magnetic field provided by a superconducting solenoid. The dc voltage and the emitted microwave power were measured simultaneously by a nanovoltmeter and a spectrum analyzer in the frequency range from 100\,MHz to 50\,GHz.
The microwave spectrometer allowed for the detection of signals with power levels down to $10^{-16}$\,W in a 25\,MHz bandwith. The spectrometer system consisted of a spectrum analyzer (Keysight Technologies N9020B, 10\,Hz -- 50\,GHz), a semirigid coaxial cable (SS304/BeCu, dc -- 61 GHz, insertion loss 6.94\,dB/m at 20\,GHz), and an ultra wide band low noise amplifier (RF-Lambda RLNA00M54GA, 0.01 -- 54\,GHz). The emitted signal was picked up by a wire loop shorting the end of a semirigid coaxial cable and placed close to the sample edge parallel to the sample surface. The diameter of the wire loop was about 2.8\,mm such that the antenna operated in a nongradient loop coupling \cite{Gol94prb} in the whole accessible frequency range. The signal was amplified by a low noise preamplifier with a gain of $36$\,dB. The signals emitted at different temperatures were further normalized by using a directional coupler with an attenuation of $-15$\,dB at $T = 1.8$\,K, $-12$\,dB at $3.0$\,K, and $-1.5$\,dB at $3.6$\,K.

\section{Acknowledgements}
OD thanks Vasyl Denysenkov for useful advices at preliminary stages of the experiment. OD acknowledges the German Research Foundation (DFG) for support through Grant No 374052683 (DO1511/3-1). This work was supported by the European Cooperation in Science and Technology via COST Action CA16218 (NANOCOHYBRI). Further, funding from the European Commission in the framework of the program Marie Sklodowska-Curie Actions --- Research and Innovation Staff Exchange (MSCA-RISE) under Grant Agreements No. 644348 (MagIC) and 645660 (TUMOCS) is acknowledged.
\vspace{2mm}

\section{Author contributions}
O.V.D. and V.A.S. conceived the experiment. M.Yu.M. and O.I.Yu. fabricated the samples. O.V.D. performed the measurements with assistance of M.I.T. and R.S.. V.M.B. and R.V.V. evaluated the emission spectra. All authors discussed the results and commented on the manuscript at all stages. O.V.D. wrote the manuscript with inputs from all authors.
\vspace{2mm}

\section{Additional information}
\textbf{Supplementary Information} accompanies this paper.

\section{Supplementary information}

\subsection{Vortex lattice configurations at matching fields}

The commensurability effect in anisotropic layered superconductors was considered theoretically by Bulaevskij and Clem (BC) \cite{Bul91prb} on the basis of the discrete Lawrence-Doniach approach and by Ivlev, Kopnin, and Pokrovsky (IKP) in the framework of the continuous Ginzburg-Landau model \cite{Ivl90ltp}. BC predicted a sequence of first-order phase transitions between vortex lattices with different matching orders at strong, parallel magnetic fields \cite{Bul91prb}. The transitions occur at $H_{n,n-1}$ expressed through the characteristic field $H_0 = \Phi_0/\gamma s^2$ at which the overlap of the Josephson cores of vortices is essential. For our sample $H_0 = 15.8$\,T such that a transition between commensurate phases with the vortex lattice period $Z_0 =s$ and $Z_0 =2s$ should occur in the field $H_{2,1} \approx H_0/3 = 5.27$\,T. Another transition between the phases with $Z_0 = 2s$ and $Z_0=3s$ is expected at $H_{3,2} \approx H_0 /8 = 1.975$\,T. While the field values calculated within the BC model corresponded well to the fields of resistance minima in superlattices with the same $d_\mathrm{Mo}=22\,\mathrm{\AA}$ but a larger $d_\mathrm{Si}=34\,\mathrm{\AA}$ with $\eta_J \approx0.7$ \cite{Mik05ltp,Mik99ltp}, the $R(H^{\parallel b})$ curve of our sample has no minima at the BC matching fields. We attribute this to a larger interlayer coupling in our sample and proceed to a comparison of the data with the continuous IKP model.

IKP showed that when the intrinsic pinning energy exceeds the elastic energy of a vortex lattice shear deformation, the vortices cannot cross the layers \cite{Ivl90ltp}. In this case the vortex lattice is always commensurate with the layered structure period $s$, and the vortex lattice period $Z_0$ is determined by the initial conditions under which the lattice was formed. Accordingly, $Z_0= Ns$, where $N$ is an integer, is independent of the external field, while the vortex lattice unit cell area varies with the field only due to vortex displacements along the layers. It was shown that the free energy of the rhombic lattice in the commensurate state as a function of $H$ has two minima corresponding to the different orientations of the unit cell vectors with respect to the layer planes. According to IKP, the conditions of stability, which correspond to the free energy minima, are $N^2s^2\gamma\sqrt{3} H^{(1)} = 2\Phi_0$ and $N^2s^2\gamma H^{(2)} = 2\sqrt{3}\Phi_0$, where the stable states of the commensurate lattices correspond to a rhombic lattice with the apex angles $\varphi^{(1)} = 2\pi/3$ and $\varphi^{(2)} =\pi/3$ in the direction of motion. In the instability region there are many metastable states corresponding to different displacements of the vortex rows relative to each other in the neighboring interlayers. These states can be dynamically accessible under the $H$ variation \cite{Lev91prl}.

The IKP matching fields in the data range are $H^{(1)}_{N=1} = 4.2$\,T and $H^{(2)}_{N=2} = 3.15$\,T, in perfect agreement with the field values at which the resistance minima are observed in Fig. 3(b) of the manuscript. Accordingly, our analysis of the resistance minima suggests that we deal with a lattice of Abrikosov rather than Josepshon vortices. At the same time, we can not rule out a crossover from Abrikosov to Josephson vortices with further decrease of the temperature, as such a crossover is known in layered systems when the Abrikosov vortex with a suppressed order parameter in its core turns into a Josephson phase vortex once its core completely fits into the insulating layer \cite{Mol12nam}. Further support in favor of dealing with Abrikosov vortices is provided by the $I$-$V$ curves allowing for a universal scaling in the flux-flow regime, which would be impossible due to a sudden dissipation reduction at the crossover from Abrikosov to Josephson vortices \cite{Mol12nam}.

\subsection{Superconductivity dimensionality crossover in the Mo/Si superlattice}

\begin{figure}[t!]
\centering
    \includegraphics[width=0.68\linewidth]{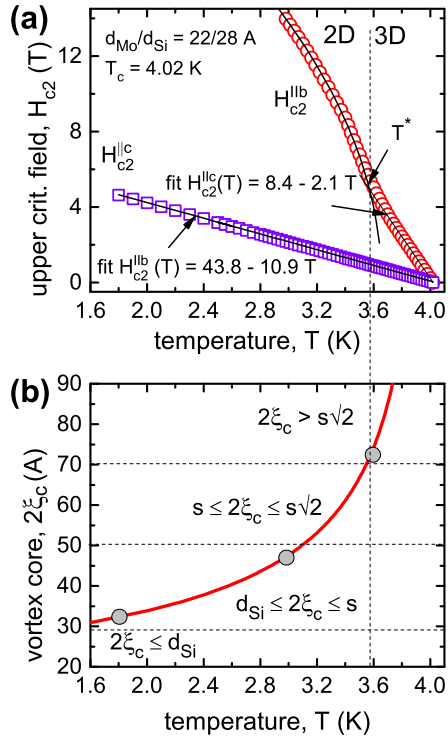}
    \caption{(a) The in-plane $H_{c2}^{\parallel b}$ and out-of-plane $H_{c2}^{\parallel c}$ upper critical fields versus temperature. Solid lines are fits $\propto(T_c-T)$ in the 3D regime and $\propto(T_c-T)^{1/2}$ in the 2D regime. The 2D-3D crossover temperature $T^\ast$ corresponding to $\xi_c(T^\ast) = s/\sqrt{2}$ is indicated. (b) Temperature dependence of the vortex core size $\simeq2\xi_c$ with the different regimes determined by the relation of $\xi_c$ and the multilayer period $s$. Large circles indicate the temperature at which the emission spectra in Fig. 2 of the manuscript have been acquired.
    }
   \label{f5}
   \vspace*{-3mm}
\end{figure}

The evolution of the matching minimum in the $R(T)$ curve at $T = 3.6$\,K to the zero-resistance state at $T = 1.8$\,K can be understood with the aid of the dimensional crossover in the Mo/Si, as inferred from the $H$-$T$ phase diagram shown in Fig. \ref{f5}(a). In Fig. \ref{f5}(a), the temperature dependence of the upper critical field $H_{c2}(T)$ is plotted for the in-plane and out-of-plane field directions. The $H_{c2}(T)$ data were deduced from the $R(T)$ curves by the $90\%$ resistance criterion. Near $T_c$, for both directions $H_{c2} \propto (1 - T/T_c)$ with slopes of $|\frac{d H_{c2}^{\parallel c}}{dT}|_{T_c} = 2.08$\,T/K and $|\frac{d H_{c2}^{\parallel b}}{dT}|_{T_c} = 10.86$\,T/K, yielding an anisotropy parameter $\gamma = 5.22$. The out-of-plane upper critical field extrapolated to zero temperature $H_{c2}^{\parallel c}(0) = 8.4$\,T yields $\xi_{ab} = [\Phi_0/2\pi H_{c2}^{\parallel c} (0)]^{1/2} = 63\,\mathrm{\AA}$ and, hence, $\xi_c(0) = \xi_{ab}(0)/\gamma = 12\,\mathrm{\AA}$. At lower temperatures $H_{c2}^{\parallel b} \propto (T_c - T)^{1/2}$, pointing to a transition at $T^\ast\approx3.6$\,K from the 3D regime of weak layering with $\xi_c(T) >70\,\mathrm{\AA}$ near $T_c$ to the 2D regime of strong layering at lower temperatures.

The increase of the size of the vortex core with increasing temperature $\simeq2\xi_c(T)$ is illustrated in Fig. \ref{f5}(b) in comparison with the thickness of the Si layer $d_{\mathrm{Si}}$ and the multilayer period $s$. A semi-quantitative relation of the vortex core size to the Si layer thickness and the multilayer period is sketched on the top of the spectra in Fig. 2 of the manuscript. In particular, at $1.8$\,K, being the lowest temperature accessible in our experiment, the vortex core $2\xi_c(1.8\,\mathrm{K}) \approx d_\mathrm{Si}=28\,\mathrm{\AA}$ largely fits into the insulating layers, thereby allowing the Mo layers to remain superconducting up to very high fields \cite{Ber97prl,Iid13prb}. At $3$\,K the vortex core $2\xi_c(3\,\mathrm{K}) \approx d_\mathrm{Si}\approx 50\,\mathrm{\AA}$ becomes comparable with the multilayer period. Even though some part of the vortices penetrates into the Mo layers, there are field ranges where the intrinsic pinning energy $E_p$ is larger than the elastic energy of a vortex lattice shear deformation $E_{el}$, which explains the presence of a rather broad resistance minimum in the vicinity of the matching fields. At $3.6$\,K the vortex cores become appreciably larger than the multilayer period, namely $2\xi_c(3.6\,\mathrm{K}) > 70\,\mathrm{\AA}$, such that the intrinsic confinement potential is smoothed out as the vortex core extends over more than one multilayer period. In this case the superlattice is no longer felt by a vortex as a layered structure, but rather the motion of vortices occurs in some effective continuous medium. Accordingly, the matching minimum at $3.15$\,T becomes shallow at $3.6$\,K while the minimum at $4.2$\,T disappears altogether as this field value is too close to $H_{c2}^{\parallel b}(\mathrm{3.6\,K}) = 5.2$\,T and it gets smeared by the transition to the normal state.


%

\end{document}